\documentclass[prl,twocolumn,showpacs,superscriptaddress]{revtex4}
\usepackage{amsmath,mathrsfs}
\usepackage{graphicx}
\begin{document}

\title{Kondo effect in a quantum dot coupled to ferromagnetic leads:
A numerical renormalization group analysis}%
\author{Mahn-Soo Choi}%
\affiliation{Department of Physics, Korea University,
Seoul 136-701 Korea}%
\author{David S\'anchez}%
\affiliation{D\'epartement de Physique Th\'eorique, Universit\'e de
  Gen\`eve, CH-1211 Gen\`eve 4, Switzerland}
\author{Rosa L\'opez}%
\affiliation{D\'epartement de Physique Th\'eorique,
Universit\'e de Gen\`eve, CH-1211 Gen\`eve 4, Switzerland}%
\date{\today}

\begin{abstract}
We investigate the effects of spin-polarized leads on the Kondo
physics of a quantum dot using the numerical renormalization group
method.  Our study demonstrates in an unambiguous way that the Kondo
effect is not necessarily suppressed by the lead polarization: While
the Kondo effect is quenched for the asymmetric Anderson model, it
survives even for finite polarizations in the regime where charge
fluctuations are negligible.  We propose the linear tunneling
magnetoresistance as an experimental signature of these behaviors.  We
also report on the influence of spin-flip processes.
\end{abstract}


\maketitle

%%%%%%
\let\eps=\epsilon%
\let\up=\uparrow%
\let\down=\downarrow%
\newcommand\varH{\mathscr{H}}%
\newcommand\half{\frac{1}{2}}%
\newcommand\ket[1]{\left|#1\right\rangle}%
\newcommand\bra[1]{\left\langle#1\right|}%
\newcommand\TMR{\mathrm{TMR}}%

%%%%%%
\emph{Introduction}.---Magnetic impurities embedded in metallic hosts
cause anomalous resonant scattering of conduction band electrons. At
the same time, the localized magnetic moments are screened at low
temperature by the itinerant electron spins.  This is the celebrated
Kondo effect~\cite{Hewson93a}, which has been recently revived in
mesoscopic physics~\cite{Kouwenhoven01a}.  Ever since the theoretical
predictions~\cite{Glazman88a,Ng88a} and the experimental
demonstrations~\cite{Goldhaber-Gordon98a}, the Kondo effect in
phase-coherent systems such as quantum dots (QD's) has stimulated
great interest in this field.  The remarkable success behind this is
the fine tunability of the parameter space (impurity level and
hybridization couplings).  The controlled manipulation in mesoscopic
systems has not only allowed to test various aspects of the Kondo
effect, which is a hard task in bulk solids, but also has posed
further exciting questions.  For example, when the
\emph{spin-degeneracy} of the impurity level is lifted by an external
magnetic field, the Kondo peak in the density of states (DOS) of the
dot is expected to split~\cite{Meir93a}. However, new
experiments~\cite{Sasaki00a} and theoretical
studies~\cite{Pustilnik00a} suggest that the situation is more subtle.

A flood of very recent works~\cite{Sergueev02a,ZhangP02a,Bulka03a,Lopez03a,
Martinek02z,Lu02z,Ma02z,Dong03z} has introduced another interesting
issue, namely, how the Kondo physics is affected when the
continuum electrons themselves are allowed to form
\emph{spin-dependent} bands.  The motivation for this research stems
from the successful field of
spintronics~\cite{Wolf01a}.  In particular, a change has been detected
in the resistivity of a Kondo alloy due to spin-polarized
currents~\cite{Taniyama03a}.  Furthermore, it is already possible to
attach ferromagnetic leads to a carbon nanotube~\cite{Tsukagoshi99a},
and a carbon-nanotube QD has been shown to display Kondo physics below
an unusually high temperature~\cite{Nygard00a}.  In addition, a QD
coupled to ferromagnetic electrodes has been proposed as a promising
candidate for spin injection devices, but so far, analyzed only in the
Coulomb blockade regime~\cite{Deshmukh02a}. The present work provides
precise theoretical predictions in a wider region of the parameter
space including the strong coupling limit.

The aim of our work is twofold. First, based upon a numerical
renormalization group (NRG) calculation, we investigate the influence
of ferromagnetic electrodes and the relative orientations of their
magnetizations on the equilibrium properties of a QD with and without
intrinsic spin flip processes.  To the best of our knowledge, this is
the first study of the model that sweeps across the different regimes
(i.e., Kondo, mixed-valence, and empty-orbital), thoroughly analyzed
with the assessment of local DOS, linear conductance, and tunneling
magnetoresistance (TMR).  Second, we resolve a controversy lately
raised in the literature with regard to whether a spin-dependent
renormalization of the impurity level induced by the spin-polarized
leads will split the Kondo peak when the magnetic moments of both
leads are aligned.  In Ref.~\cite{Sergueev02a}, an equation-of-motion
(EOM) method plus an ansatz for the interacting
self-energy~\cite{Ng93a} were employed and it was suggested that the
splitting, $\delta$, is absent.  In a later work~\cite{Martinek02z},
the scaling arguments (together with the EOM method) were used to find
that $\delta$ is nonzero.  In
Ref.~\cite{ZhangP02a}, using a similar approach, a splitting was
predicted only in the mean-field peaks.  In a more recent
work~\cite{Dong03z}, they made use of a noncrossing approximation
(NCA) and obtained $\delta\neq 0$.  In
Refs.~\cite{Bulka03a,Lopez03a}, on the other hand, the slave-boson
mean-field theory (SBMFT) was utilized to study the zero-temperature
properties and no splitting was observed. The answer to the
controversy is, thus, elusive because each approximation method
mentioned above has certain drawbacks of its own.  Below, according to
a NRG calculation, which has been known to provide very accurate
results for impurity problems~\cite{Krishna80a}, we resolve clearly
the controversy.

\begin{figure}
\centering%
\includegraphics*[width=0.47\textwidth,angle=0,clip]{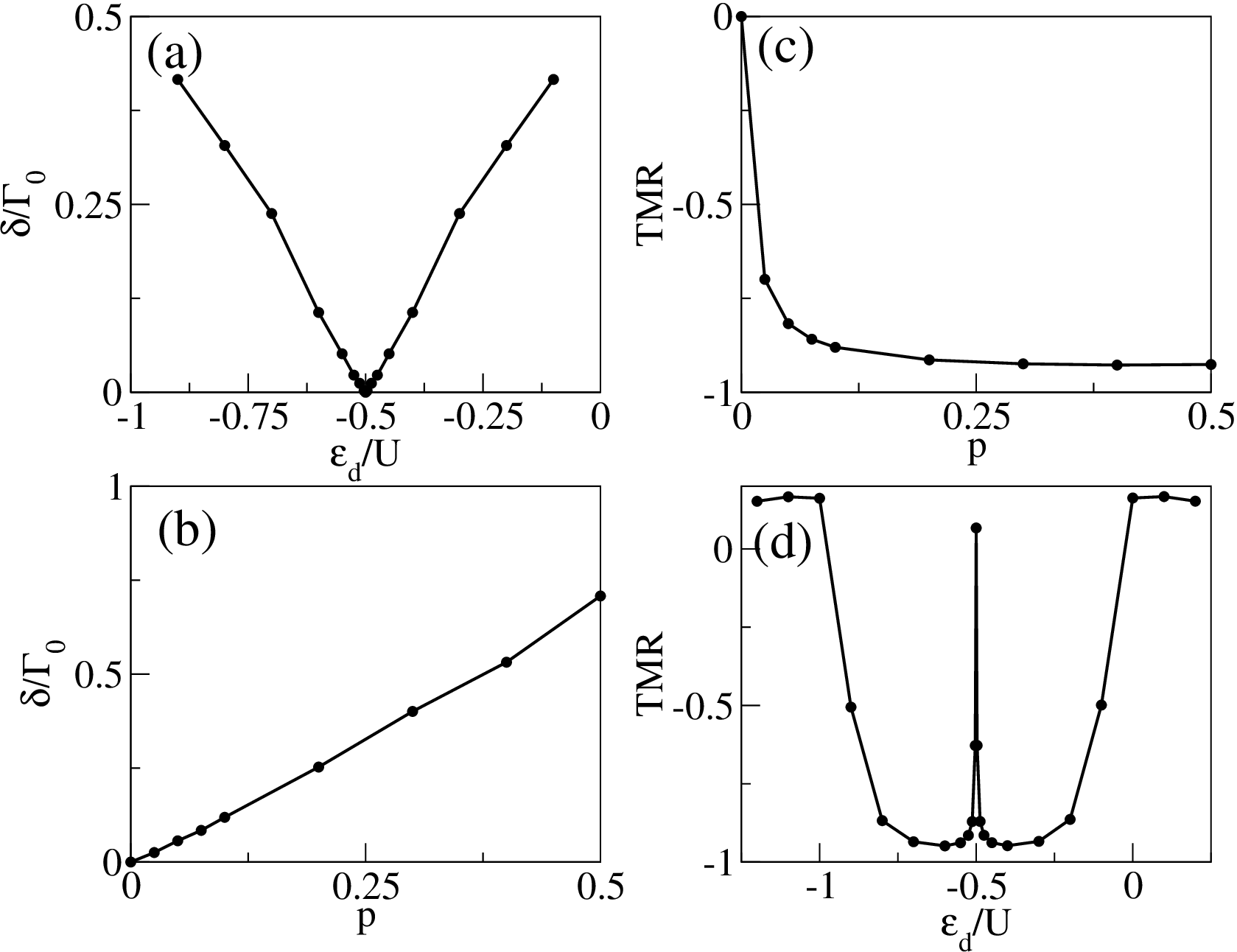}
\caption{(a) Splitting $\delta$ of the Kondo peak
  as a function of $\varepsilon_d$ for $p=0.25$ and $U=0.4D$.  (b)
  $\delta$ versus $p$ for $\varepsilon_d=-0.1D$ and $U=0.45D$. (c) TMR
  versus $p$ for $\varepsilon_d=-0.1D$ and $U=0.4D$. (d) TMR as a
  function of $\varepsilon_d$ for $p=0.25$ and $U=0.4D$. In all cases
  $\Gamma_0=0.02D$.}
\label{paper::fig:2}
\end{figure}

Our main result is summarized in Fig.~\ref{paper::fig:2}(a).  We find
that in the presence of electron-hole symmetry the Kondo peak at the
Fermi level remains unsplit even at finite polarizations and the
linear conductance achieves the unitary limit. This remains
true as long as only spin fluctuations are present in the QD.  On the
contrary, when charge fluctuations start to play a role (as in the
asymmetric case of the Anderson model), the Kondo peak shows a visible
splitting and the conductance is then suppressed.

\begin{figure}
\centering%
\includegraphics*[width=0.47\textwidth,angle=0,clip]{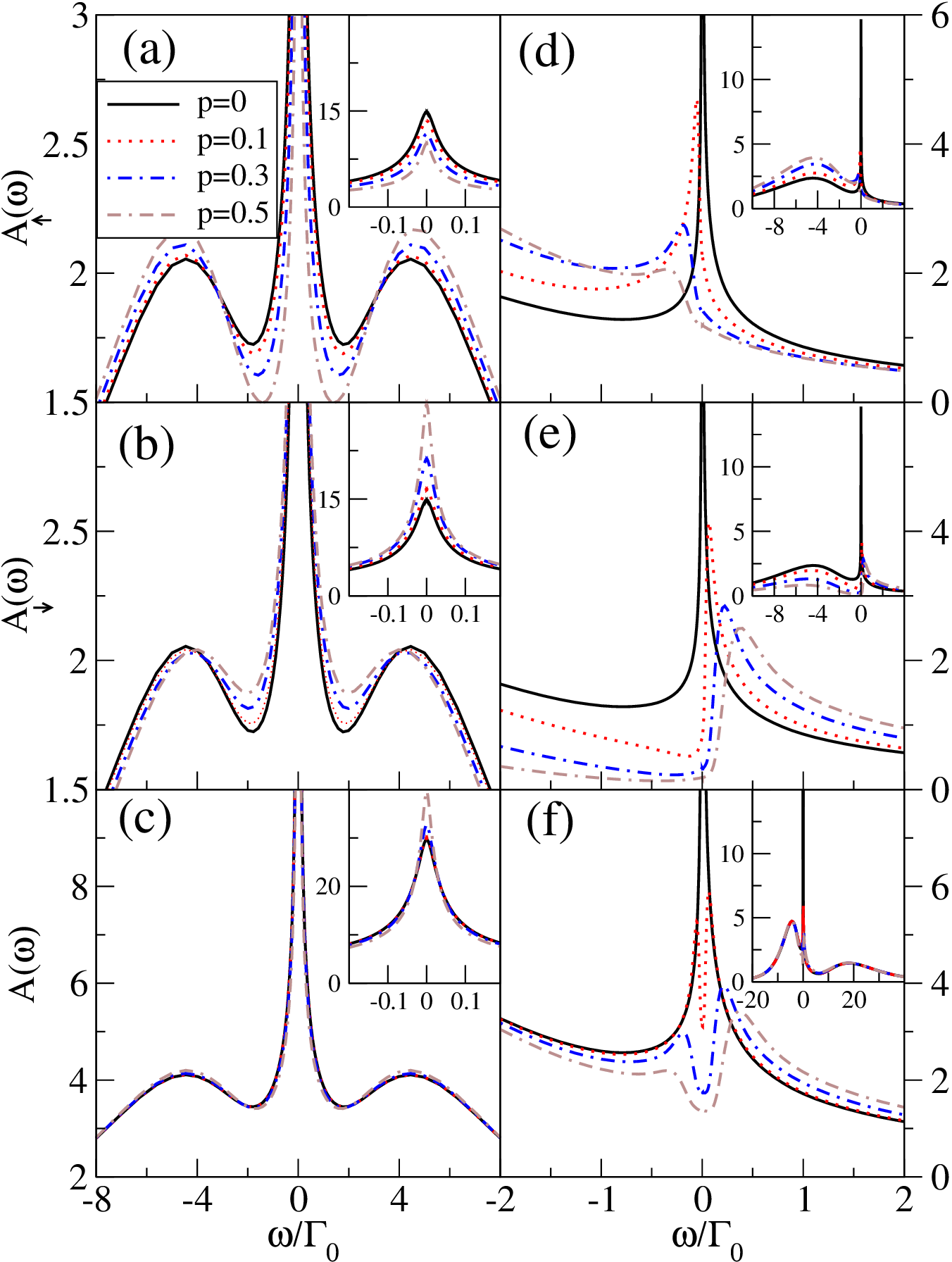}
\caption{Left: Local DOS of the QD for the symmetric Anderson model. (a)
  $A_\up(\omega)$, (b) $A_\down(\omega)$, and (c) $A(\omega)$ for
  $\varepsilon_d=-U/2=-0.1D$ ($D$ is the bandwidth).
  Right: DOS for the asymmetric case. (d) $A_\up(\omega)$, (e)
  $A_\down(\omega)$, and (f) $A(\omega)$ for $\varepsilon_d=-0.1D$ and
  $U=0.45D$. In all cases $\Gamma_0=0.02D$.}
\label{paper::fig:1}
\end{figure}

\emph{Model}.---We consider an ultrasmall tunnel junction comprising a
QD coupled to two ferromagnetic leads.  We assume that the QD consists
of an energy level with an unpaired spin-$1/2$ electron and a charging
energy $U$.  This way, the QD is equivalent to an Anderson-type
impurity with single-particle energy
$\varepsilon_{d,\sigma}$ for spin
$\sigma=\{\up,\down\}$~\cite{Glazman88a,Ng88a}.  Notice that
$\varepsilon_{d,\sigma}$ includes the Zeeman
energy $\Delta_Z\equiv\varepsilon_{d,\up}-\varepsilon_{d,\down}$ of an
external magnetic field.  In what follows, we set $\Delta_Z=0$ in
order to unmask possible spin-dependent renormalizations of the bare
energy level purely due to coupling with the leads~\cite{note1}.  On
the other hand, we shall include in a phenomenological way internal
spin-flip scattering processes with rates $2R/\hbar$~\cite{note2}.
Tunneling of electrons from the QD to the leads (reservoirs)
$\alpha=\{L,R\}$ is described by the hopping integral $V_{\alpha
  k\sigma}$.  The resulting Hamiltonian is given by:
\begin{multline}
\label{paper::eq:H}
\varH = \sum_{\sigma}\varepsilon_{d,\sigma}\hat{n}_\sigma +U
\hat{n}_\uparrow \hat{n}_\downarrow
+ R\left(d_\up^\dag d_\down + {\rm H.c.}\right) \\
+ \sum_{\alpha k\sigma}\varepsilon_{\alpha k\sigma} c_{\alpha
  k\sigma}^\dagger c_{\alpha k\sigma} + \sum_{\alpha k\sigma}
(V_{\alpha k\sigma} c_{\alpha k\sigma}^\dagger d_{\sigma}+ {\rm H.c.})
\,,
\end{multline}
where $c_{\alpha k\sigma}^\dagger$ ($c_{\alpha k\sigma}$) is the
creation (annihilation) operator for an electron with wave vector $k$
and spin in the electrode $\alpha$. The QD occupation number is
$\hat{n}_\sigma=d_{\sigma}^\dagger d_{\sigma}$ [$d_{\sigma}^\dagger$
($d_{\sigma}$) creates (annihilates) an electron in the dot].

For definiteness, we shall take identical leads with chemical
potentials $\mu_L=\mu_R=E_F$ and symmetric couplings (i.e., equal
tunnel barriers).  Ferromagnetism on the leads may be represented
either by a spin-dependent DOS $\rho_{\alpha\sigma}(\omega)$ or by
spin-dependent tunneling amplitudes $V_{\alpha k\sigma}$.  We choose
the latter for convenience but both pictures are formally equivalent
provided we are interested just in the transport properties of the
junction.  In any case, the overall effect results in a spin-dependent
hybridization parameter $\Gamma_{\alpha\sigma}(\omega)\equiv \pi
\sum_k |V_{\alpha k\sigma}|^2\delta(\omega-\varepsilon_{\alpha k})$.
(We neglect proximity effects such as stray fields induced in the QD).
A further simplification consists of neglecting the energy dependence
of $\Gamma_{\alpha\sigma}(\omega)$, evaluating it at $\omega=E_F$
(wide band limit).
In the following, we choose $E_F=0$ as the origin of energies.
Therefore, one could define the spin polarization (close to the Fermi
energy) at each lead as $p_\alpha =
(\Gamma_{\alpha\up}-\Gamma_{\alpha\down})/
(\Gamma_{\alpha\up}+\Gamma_{\alpha\down})$ with $-1\leq p_\alpha \leq
1$.  We consider parallel (P) and antiparallel (AP) magnetizations of
the two leads. In the P case ($p_L=p_R\equiv{p}$), we have
$\Gamma_{L\up} = \Gamma_{R\up} = (1+p)\Gamma_0/2$ and $\Gamma_{L\down}
= \Gamma_{R\down} = (1-p)\Gamma_0/2$, where
$\Gamma_0\equiv\Gamma_{\alpha\up}+\Gamma_{\alpha\down}$, whereas the
AP case ($p_L=-p_R\equiv{p}$) yields $\Gamma_{L\up} = \Gamma_{R\down}
= (1+p)\Gamma_0/2$ and $\Gamma_{L\down} = \Gamma_{R\up} =
(1-p)\Gamma_0/2$.

In order to apply the NRG technique more efficiently, we map
Eq.~(\ref{paper::eq:H}) onto an effective model with a \emph{single}
lead. This is achieved by means of a canonical
transformation~\cite{Glazman88a,Buttiker88a}. In the P configuration
it reads (we omit the index $k$ for simplifying notation):
\begin{equation}
\label{paper::eq:P}
c_{L\sigma}= (a_{\sigma}+b_{\sigma})/\sqrt{2} \,,\quad
c_{R\sigma}= (a_{\sigma}-b_{\sigma})/\sqrt{2} \,.
\end{equation}
For AP polarizations one uses the relations:
\begin{subequations}
\label{paper::eq:AP}
\begin{align}
c_{L\up}&= (V_{a}a_{\up}+V_{i}b_{\up})/\mathcal{V}\,, \,\,\,\,
c_{R\up}= (V_{i}a_{\up}-V_{a}b_{\down})/\mathcal{V} \,, \\
c_{L\down}&= (V_{i}a_{\down}+V_{a}b_{\down})/\mathcal{V} \,,\,\,\,\,
c_{R\down}= (V_{a}a_{\down}-V_{i}b_{\down})/\mathcal{V} \,,
\end{align}
\end{subequations}
where $V_{a}=V_{L\up}=V_{R\down}$ ($V_{i}=V_{L\down}=V_{R\up}$) is the
tunneling amplitude for m\emph{a}jority (m\emph{i}nority) spins and
$\mathcal{V}=\sqrt{|V_{a}|^2+|V_{i}|^2}$.  Substituting
Eqs.~(\ref{paper::eq:P})--(\ref{paper::eq:AP}) into
Eq.~(\ref{paper::eq:H}) one can show that (i) the QD electron
decouples from the $b_{\sigma}$ operators and hybridizes only with the
quasiparticles described by the $a_{\sigma}$ operators; and (ii) the
effective dot-lead couplings are renormalized for both configurations.

We use the NRG method to obtain the local DOS
$A_\sigma(\omega)=A_\sigma^>(\omega)-A_\sigma^<(\omega)$ on the QD
(i.e., at the impurity site).  The particle and hole branches,
$A_\sigma^>(\omega)$ and $A_\sigma^<(\omega)$, are defined by
\begin{equation}\label{paper::eq:DOS}
A_\sigma^{>/<} (\omega) = \pm\sum_m
\left|\bra{m}d_\sigma^\dag /d_\sigma \ket{0} \right|^2
\delta(\omega\mp E_m\pm E_0) \,,
\end{equation}
%\begin{subequations}
%\label{paper::eq:DOS}
%\begin{align}
%A_\sigma^>(\omega) &= \sum_m
%\left|\bra{m}d_\sigma^\dag\ket{0}\right|^2
%\delta(\omega-E_m+E_0) \,, \\
%A_\sigma^<(\omega) &= -\sum_m \left|\bra{m}d_\sigma\ket{0}\right|^2
%\delta(\omega+E_m-E_0) \,,
%\end{align}
%\end{subequations}
%respectively,
where $\ket{0}$ is the many-body ground state
(electrodes + QD) and $\ket{m}$ is an excited state ($E_0$ and $E_m$
are the corresponding energies). Notice that all the physics
(correlations, dependence on the gate voltage, etc.) is contained in
$A_\sigma(\omega)$.  The linear conductance (normalized to
$e^2/h$) of the junction at zero temperature is obtained from the
impurity spectral density function at the Fermi level~\cite{Meir92a},
$g=\sum_\sigma 2\Gamma_{L\sigma} \Gamma_{R\sigma} A_\sigma (0)/
(\Gamma_{L\sigma}+\Gamma_{R\sigma})$.  As it is the case in usual
Kondo experiments on QD's, we shall present results for large
interaction ($U\gg\Gamma$).

\emph{Results}.---We first address the issue whether the Kondo peak in
local DOS splits or not. (Until later, we put $R=0$.)
Figure~\ref{paper::fig:1} shows $A_\up(\omega)$,
$A_\down(\omega)$, and $A(\omega)=A_\up(\omega)+A_\down(\omega)$ for
different values of the lead polarization $p$ in the P configuration.
[The AP case is less interesting as both spin orientations are equally
coupled after the transformation given by Eq.~(\ref{paper::eq:AP})].
Left panels correspond to the symmetric Anderson model (i.e.,
$\varepsilon_d=-U/2$).  For $p=0$, in addition to two (symmetric)
mean-field peaks at $\omega=\varepsilon_d$ and
$\omega=\varepsilon_d+U$, $A(\omega)$ shows a peak at $\omega=0$ [see
Fig.~\ref{paper::fig:1}(c)], which is responsible for the observed
zero-bias anomaly.  As $p$ increases, the spectral peak of
$A_\up(\omega)$ [$A_\down(\omega)$] at the Fermi energy increases
(decreases) [see insets of Fig.~\ref{paper::fig:1}(b) and (c)].
Remarkably, however, the central peaks of both $A_\up(\omega)$ and
$A_\down(\omega)$ are pinned at the Fermi level; in particular, the
Kondo peak in $A(\omega)$ does \emph{not} split. Experimentally, one
would see a perfect transparency of the junction (see below). Right
panels of Fig.~\ref{paper::fig:1} show the same functions for the
asymmetric case ($\varepsilon_d\neq{}-U/2$), where charge fluctuations
are allowed to certain extent.  As $p$ increases, $A_\up$ and
$A_\down$ shift in opposite directions [see Fig.~\ref{paper::fig:1}(d)
and (e)] and the Kondo peak in $A(\omega)$ splits into two
[Fig.~\ref{paper::fig:1}(f)].  As a result, the Kondo effect is
suppressed.  We have checked as well that both mean-field peaks are
shifted in opposite directions, though their strengths differ and the
splitting cannot be resolved in
Fig.~\ref{paper::fig:1}(f). 
\begin{figure}
\centering%
\includegraphics*[width=0.47\textwidth,angle=0,clip]{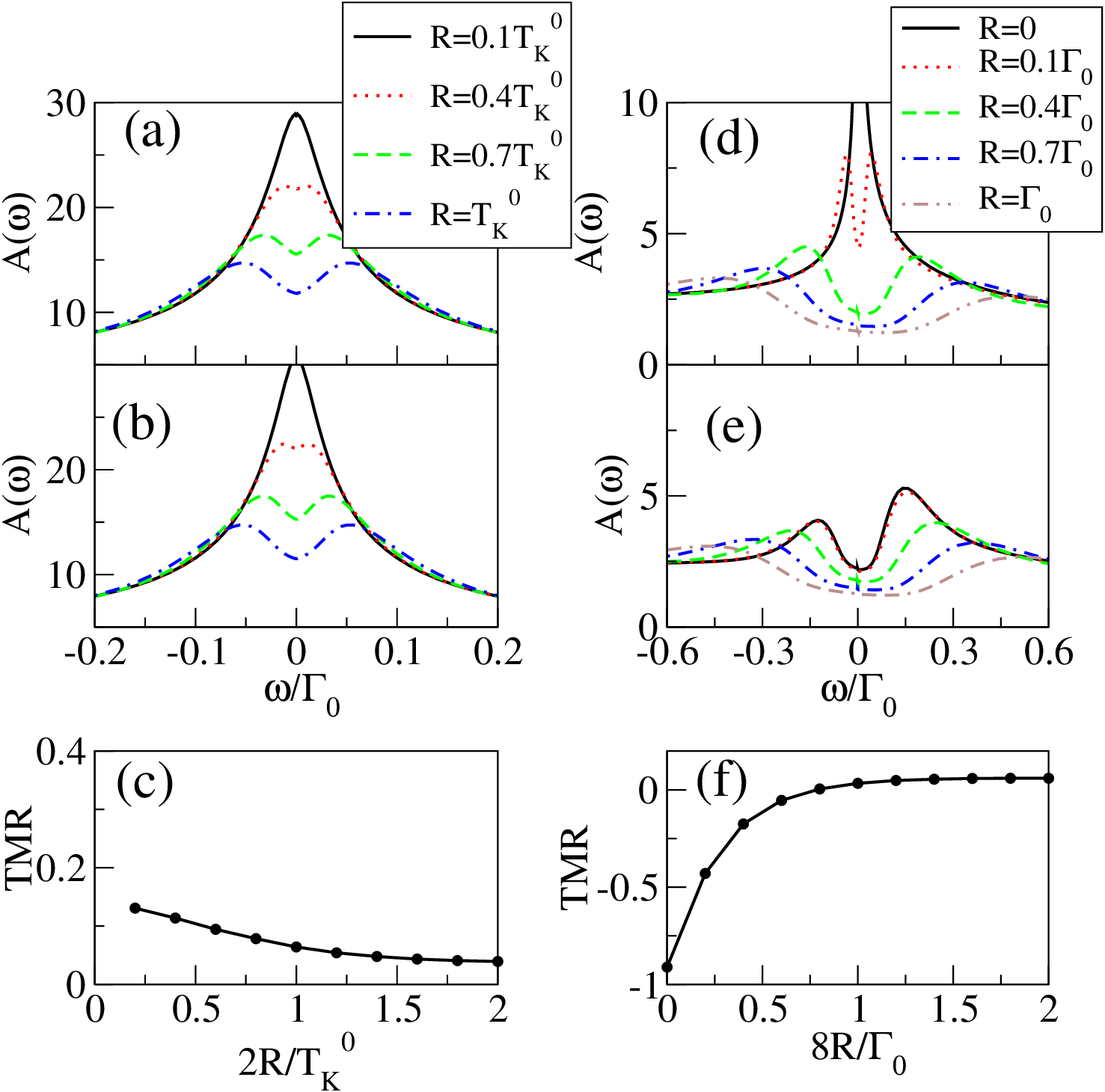}
\caption{The effects of spin-flip processes $R$ for the symmetric
  (left panels with $\varepsilon_d=-U/2=-0.1D$) and asymmetric (right
  panels with $\varepsilon_d=-0.1D$ and $U=0.4D$) Anderson model. Local
  DOS for (a,d) $p=0$ and (b,e) $p=0.25$.  (c,f) TMR versus $R$ for
  $p=0.25$.  In all cases $\Gamma_0=0.02D$.}
\label{paper::fig:3}
\end{figure}

The finite splitting for the asymmetric Anderson model may be understood
in terms of simple scaling arguments~\cite{Martinek02z}: Because the
hybridization for up spins is larger than for down spins
($\Gamma_{\alpha\up}>\Gamma_{\alpha\down}$), the renormalization of the
bare level $\varepsilon_d$ is spin-dependent; the
$\up$($\down$)-electron lowers (raises) its energy.  Then, the
coupling acts as an effective magnetic field, leading to a finite
$\delta$~\cite{ChoiMS03c}.
Yet, the perturbative nature of a poor man's scaling cannot describe the
fixed point in the strong coupling regime.  In particular, such simple
scaling arguments cannot account properly for the particle-hole symmetry
in the symmetric Anderson model and always predict $\delta\neq 0$.
For the symmetric Anderson model, it is important to notice that the
particle-hole symmetry quenches charge fluctuations completely for both
spins ($\langle n_\up \rangle=\langle n_\down \rangle=1/2$) at \emph{any}
$|p|<1$, and the real part of the self-energy (at $E_F$) is zero.
This means that although the binding energy of the singlet state
(the Kondo temperature $T_K$) diminishes with $p$, the quasiparticle
lifetime is still infinite and the Fermi liquid picture is valid. 
Therefore, the results in Fig.~\ref{paper::fig:1}(c) are
consistent with SBMFT, which describes the Kondo peak when spin
fluctuations prevail. Likewise, the results in
Fig.~\ref{paper::fig:1}(f) are in agreement with EOM and NCA models,
which support charge fluctuations to some degree.  Of course, the
NRG method can encompass the whole parameter range.

To illustrate our conclusions, we measure the splitting $\delta$ of the Kondo
peak as a function of $\varepsilon_d$ (experimentally this is
controlled by a gate voltage) with $U$ fixed [see
Fig.~\ref{paper::fig:2}(a)]. The splitting $\delta$
increases roughly linearly from zero as moving away from the symmetric
point $\varepsilon_d=-U/2$.  Notice also that well away from
$\varepsilon_d=-U/2$, $\delta$ is linear in the lead polarization [see
Fig.~\ref{paper::fig:2}(b)], confirming the prediction relying upon
scaling arguments.

We now turn to the tunneling magnetoresistance of the system, defined
by $\TMR = (g^{P}-g^{AP})/g^{AP}$. One can show that $g^{P} =
\pi\Gamma_0\left[(1+p)A_\up(0) + (1-p)A_\down(0)\right]$ and $g^{AP} =
(1-p^2)\pi\Gamma_0 A(0)$ are the dimensionless linear conductances for
the P and AP configurations, respectively.  For the symmetric Anderson
model, the Kondo effect survives even for a finite value of
polarization $p$ ($|p|<1$), and $g^{P}$ preserves the unitary limit.
As a result, the TMR is given by $\TMR=p^2/(1-p^2)$.  For
the asymmetric Anderson model, on the contrary, $g^{P}$ gets strongly
suppressed as $p$ increases. Then, the system exhibits a
strong negative TMR [see Fig.~\ref{paper::fig:2}(c)]~\cite{note4}.
Figure~\ref{paper::fig:2}(d) shows TMR as a function of
$\varepsilon_d$,
which shows a sharp peak around the symmetry point
($\varepsilon_d\simeq -U/2$).
The width of the peak is determined by how fast the Kondo effect is
suppressed as $|\varepsilon_d-U/2|$ increases from zero, and hence
depends strongly on the polarization $p$ and the hybridization
$\Gamma_0$; see Figs.~\ref{paper::fig:2} (a) and (b).  Experimentally,
finite temperatures would smoothen this peak.

So far we ignored the spin-flip scattering.
$R\neq 0$ introduces a new degree of freedom in the QD, explicitly
lifting the level degeneracy ($\varepsilon_d\pm R$).  In
Fig.~\ref{paper::fig:3}(a,b) we plot the local DOS in the symmetric
case for different values of $R$ in terms of the Kondo temperature
$k_B T_K^0 = \sqrt{\Gamma_0 U/2} \exp
[-\pi|\varepsilon_d(\varepsilon_d+U)|/2\Gamma_0 U]$.
For $p=0$ the Kondo peak
shrinks~\cite{note3} and splits as expected (the effect can be
ascribed to an external Zeeman field).  For $p=0.25$ the splitting
takes place for a smaller value of $R/T_K^0$. As a result, $T_K$ in
the P case is always lesser than $T_K^0$ and TMR decreases in
magnitude with increasing $R$ [see Fig.~\ref{paper::fig:3}(c)].  The
DOS for the asymmetric Anderson model are shown in
Fig.~\ref{paper::fig:3}(d,e) as a function of $R/\Gamma_0$ as here we
are interested in the competition between both splitting factors ($p$
and $R$). For a nonzero value of $p$, the splitting is not symmetric
unlike Fig.~\ref{paper::fig:3}(b).  This is again a consequence of the
presence of charge fluctuations.  For $R=0$ the DOS in the P alignment
is already quenched due to the spin dependent coupling, which leads to
a strong negative TMR.  However, this effect is washed out with
increasing $R$ and TMR tends to vanish [see
Fig.~\ref{paper::fig:3}(f)].

\emph{Conclusion}.---Using a NRG method,
we have shown that the Kondo effect in a quantum dot is not
necessarily suppressed by the spin polarizations of the leads: for the
symmetric Anderson model, where charge fluctuations are completely
suppressed, the Kondo effect is robust even for finite polarizations.
For the asymmetric Anderson model, the Kondo peak does split into two.
We also reported on the TMR in the Kondo, mixed-valence, and
empty-level regimes, being strongly affected in the presence of spin
flip processes.  The physics addressed in this paper is realistic and
can be, in principle, tested with present techniques.

We thank M.~B\"uttiker, T.~Costi and K.~Kang for valuable comments.
M.-S.C.\ acknowledges supports from the
SKORE-A program, from the eSSC at POSTECH, and from a Korea University
Grant.  D.S. and R.L. were supported by the Swiss NSF through the
program MANEP, and by the Spanish MECD.

\emph{Note Added}.---In the final stages of this work we became aware of
a closely related work by J.~Martinek \textit{et al.}, cond-mat/0304385.
The difference is that they study only the asymmetric Anderson model,
and focus rather on the restoring effect of an external magnetic field.

%%%%%% References
% \bibliographystyle{physrev}
% \bibliography{aliases,cond-mat,staphy,choims,paper}

\begin{thebibliography}{10}
\bibitem{note1} The contribution from finite $\Delta_Z$ to the
  spin-dependent Kondo peak is (almost) independent of $\delta$ and can
  be used to compensate the latter; see Refs.~\cite{Martinek02z} and
  \cite{ChoiMS03c}.
\bibitem{note4}%
  This should not be confused with the behavior of tunnel junctions with
  a \emph{noninteracting} resonant level, where one usually has
  $g^{P}>g^{AP}$ and TMR increases with $p$ [see, e.g., A.M.
  Bratkovsky, Phys. Rev. B \textbf{56}, 2344 (1997)].
\end{thebibliography}

\end{document}